\documentclass[aps,superscriptaddress,nofootinbib]{revtex4}
\usepackage{graphicx}
\newcommand{\non}{ \nonumber \\}
\newcommand{\be}{\begin{equation}}
\newcommand{\ee}{\end{equation}}

\begin{document}
\title{Delayed information flow effect in economy systems. \\ An ACP model study. }
\author{Janusz Mi\'s{}kiewicz}
\email{jamis@ift.uni.wroc.pl}
\affiliation{Institute of Theoretical Physics, University of Wroc\l{}aw, pl. M.
Borna 9, 50-204 Wroc\l{}aw, Poland}
\author{M. Ausloos}
\email{marcel.ausloos@ulg.ac.be}
\affiliation{SUPRATECS, B5, University of Li$\grave e$ge, B-4000 Li$\grave e$ge, Euroland}

\begin{abstract}
Applying any strategy requires some knowledge about the past state of the system. Unfortunately in the case of economy, collecting information is a difficult, expensive and time consuming process. Therefore the information about the system is usually known only at the end of some well defined intervals, e. g. through company, national bank inflation data and Gross Domestic Product (GDP) reports, etc. They describe a (market) situation in the past. The time delay is specific to the market branch. It can be very short (e.g. stock market offer is updated every minute or so and this information is quasi immediately available) or long, like months in the case of agricultural markets, when the decisions are taken based on the results from the previous harvest.

The analysis of the information flow delay can be based on the Ausloos-Clippe-P\c{e}kalski (ACP) model of spatial evolution of economic systems. The entities can move on a (square) lattice and when meeting take one of the two following decisions: merge or create a new entity. The decision is based on the system state, which is known with some time delay. The effect of system's feedback is hereby investigated. We consider the case of company distribution evolution in a heterogeneous field. The information flow time delay implies different final states, including cycles; it is like a control parameter in a logistic map.

\end{abstract}


\maketitle

\section{Introduction}

Economy system consists of companies which are working in some environment. All firms need some resources to be able to perform their job. Besides typical resources such as energy, water or labour forces there is one which plays a special role. This is the information about the system. This is one of the key points in an economy, because the information about the system state and actions of the competiting companies, allows to build appropriate strategies on the market \cite{games1,games2,games3}.
On the other hand it is impossible to know the state of the system at once. Any of economic index, e.g. Gross Domestic Product, inflation rate, ... is known with some time delay. Even in the case of the stock market one knows only the last price, based on the bids from the last session, but one does not know the bids on the present session. So even in this case the state of the system is not wholly known. The problem is best seen in the case of agriculture markets \cite{MA1}. 
Because farmers make their decisions based on the informations of the income of the latest harvest, it means that there might be a one year time delay.  It is well known that the agricultural market without control and the more so with control \cite{vitanov} is evolving as a periodic function, the typical mechanism going as follows. If there is a great demand for a product, then the prices are high. This results in enhancing production process, but usually too many farmers have the same idea which results in a great supply, which finally leads to price falling. Of course nobody wants to loose money so farmers decide to change their product. Then supply falls and prices grow, repeating the cycle.
Therefore the delay of information flow is one of the basic features of the economy system. This paper deals with the problem of information flow onto the evolution of the economy system, suggesting that the time delay is a control parameter.

The economy environment is modelled by adopting the ACP model \cite{ACP3,ACP2,ACP1}. The ACP model is altered by introducing a strategy depending on the state of the system. The influence of the information flow delay onto the behaviour of the system was investigated in the case of uniform distribution of companies in the system in \cite{ACP_AM3}. The following step is to consider the case of a non uniform distribution of firms.

The paper is organised as follows: a short description of the  ACP model is presented in Sec.\ref{ACP}, then the results of a mean field approximation are recalled in Sec.\ref{uniform}. The results of heterogeneous space evolution is presented in Sec.\ref{heterogenous}.

\section{ACP model}
\label{ACP}

For the sake of clarity a short description of the main features of the ACP model is presented below.
The model was built to simulate the behaviour of economic system in changing conditions e.g. political changes or destruction of economy barriers \cite{ACP3,ACP2,ACP1}. Originally the model was set in the form of a Monte Carlo simulation.\\

Economy entities are initially randomly placed on a square symmetry lattice. The environment is characterised by a real field $F \in [0,1]$ and a selection pressure $sel$. Each company is described by a real parameter $f \in [0,1]$ named the fitness.

 There are three possible events in the system:
\begin{enumerate}
\item the company survive with the probability $ p_i = \exp (-sel |f_i -F | )$
\item then they move on the lattice (horizontally or vertically) one step at a time if a free space is available
\item if companies meet they can set a new firm or merge; the decision is taken through the strategy parameter $b$, which is the probability of merging. Since the creation of company is a complementary event the settlement probability is $1-b$.
\end{enumerate}
The model was rewritten in the case of a mean field approximation \cite{ACP_AM1,ACP_AM2} by introducing the company distribution function $N(t,f)$, which is the number of firms with fitness $f$ at time $t$. Then the system was characterised by the concentration of companies $c(t)$ and investigated in the case of the best adapted companies $f=F$ \cite{ACP_AM1}.

The ACP model was investigated by Monte Carlo simulations in order to describe the adaptation process to new conditions \cite{ACP3,ACP2,ACP1}. Since the results show that usually the most adapted ones survive the best adapted case was investigated in order to find out stability conditions of the system \cite{ACP_AM1}.

\section{The information transfer in a uniformly occupied space}
\label{uniform}

The information transfer delay was investigated by the introduction of an investment strategy depending on the state of the system the system. The constant probability of company creation was replaced by $ b \rightarrow (1-c)$, where c was the concentration updated every n-th iteration step.
The evolution of the system was governed by the following equation:
\be
\label{evol}
c_t = c_{t-1} + \frac{1}{2} c_{t-1}(1-c^8_{t-1})(1-(1-c_{t-1})^8) (2 ST(c(g(t)))-1),
\ee
where
$ST(c) = 1-c$, $g(t) = k [\frac{t}{k}]$ and $[ \, ]$  denotes the procedure of
taking a natural number not larger than the one given in the brackets.
The time is measured in iteration steps $IS$.
It has been shown \cite{ACP_AM3} that for 
\begin{itemize}
\item short time delay, $t_d \in [2,4]$: \\ 
The system evolves to a stable state of concentration $c=0.5$. The stability time $(t_s)$ defined as the time required by the system to achieve a stable state increases with the time delay from about $t_s \approx 20 \; IS$ for $t_d \approx 2 \;IS$, $t_s=100 \; IS$ for $t_d=3 \;IS$ and $t_s \approx 4200 \; IS$ for $t_d=4 \; IS$.
\item medium time delay, $t_d=5 \; IS$ or $t_d=6 \; IS$: \\ 
Due to the time delay, cycles become visible in the system. For $t_d = 5 \; IS$ the system evolves between six states.
\item long time delay, $t_d \geq 7 \; IS$: \\
For the $t_d= 7 \; IS$, crashes appear in the system. A crash of the system is recognised if the concentration of companies goes below zero. However for a time delay longer than 7 $IS$ the system does not have to crash. Even in the case of very long time delay (12 $IS$ or 15 $IS$) and for some initial values of the concentration (e.g. $c_0 = 0.041$)  the crash time occurs after a very long time (above 400 $IS$). It has been observed that in the case of $t_d=15 \; IS$ for a close by initial concentration (e.g. $c_0 = 0.021$) the system does not crash but reaches the stable fully occupied state $(c=1)$.
\end{itemize}

\section{The information transfer a non-uniformly occupied space}
\label{heterogenous}

After the analysis of the behaviour of a uniform distribution of companies, let us consider the problem of a heterogeneous distribution of companies. The system is defined as follows:\\
\begin{itemize}
\item the space is a square symmetry lattice with a periodic boundary conditions
\item the lattice consists of subspaces described by a uniform density of firms
\item the companies may:
\begin{itemize}
\item stay at their subspace
\item jump to another lattice subspace
\item after moving (or staying)   may create a new one or merge with another
\end{itemize}
\end{itemize}
Strategy constraints:
\begin{itemize}
\item the length of the jump depends on the company strategy $J$ and the concentration of firms near the considered lattice point; it is assumed that a high  concentration causes the need for a longer jump, in order to change the place significantly
\item the decision upon creation or annihilation depends on the concentration of companies  at the new place (after the jump decision)
\item the information time delay influences only the decision about company creation.
\end{itemize}
It is stressed that the company should know the situation in its closest neighbourhood so that the decision about jump length is taken on the real data, but the creation decision depends on a global situation on the system based on the information published by the system. However the system updates the information every $t_d \; IS$, whence the decision is taken based on the information which can be as old as $t_d \; IS$.

The evolution of the system is described by the following set of equations for the increase (decrease) of concentration at point $(x,y) $ at the given time $t$ under a delay $t_d$. The eight possible directions of jumps are considered as well as the case when no jump occurs, i.e.,  $\Delta c_i (x_j,y_k)(t) $, $i=1...9$, $ j,k = 0,1,2$:\\

\begin{eqnarray}
\label{evol_new}
\Delta c_1 (x,y)(t) & \equiv & c(x,y)(t-1) \{ (1 - c(x,y)(t-1)) \cdot (1 -  c(x,y)(t_d)) \cdot (1 - c^8(x,y)(t))  - \non
& & [1 - c(x,y)(t-1)] \cdot c(x,y)(t_d) \cdot 
 [1-(1-c(x,y)(t-1))^8)] \} 
\end{eqnarray}

\begin{eqnarray}
\label{evol1}
\Delta c_2 (x_1,  y_1) (t) & \equiv & \frac{1}{8} \cdot c(x, y)(t-1) \{ (1-c(x_1, y_1)(t_d) \cdot (1-c^8(x_1, y_1)(t -1)) - \non
& & c(x_1, y_1)(t_d) \cdot [1-(1-c(x_1, y_1)(t-1))^8)] \}
\end{eqnarray}

\begin{eqnarray}
\label{evol2}
\Delta c_3(x_1, y_2) (t) & \equiv & \frac{1}{8} \cdot c(x, y)(t-1)  \{ (1-c(x_1, y_2)(t_d)) \cdot (1-c^8(x_1, y_2)(t -1)) - \non
& & c(x_1, y_2)(t_d) \cdot [1-(1-c(x_1, y_2)(t-1))^8)] \}
\end{eqnarray}

\begin{eqnarray}
\label{evol3}
\Delta c_4(x_2, y_1) (t) & \equiv & \frac{1}{8} \cdot c(x, y)(t-1)  \{ (1-c(x_2, y_1)(t_d)) \cdot (1- c^8(x_2, y_1)(t-1)) - \non
& & c(x_2, y_1)(t_d) \cdot [1-(1-c(x_2, y_1)(t-1))^8)] \}
\end{eqnarray}
\newpage
\begin{eqnarray}
\label{evol4}
\Delta c_5(x_2, y_2)(t) & \equiv &  \frac{1}{8} \cdot c(x, y)(t-1)  \{ ((1-c(x_2, y_2)(t_d)) \cdot (1-c^8(x_2, y_2)(t-1)) - \non 
& & c(x_2, y_2)(t_d) \cdot [1-(1-c(x_2, y_2)(t-1))^8)] \}
\end{eqnarray}

\begin{eqnarray}
\Delta c_6(x_1, y)(t) & \equiv & \frac{1}{8} \cdot c(x, y)(t-1)  \{ ((1-c(x_1, y)(t_d)) \cdot (1-c^8(x_1, y)( t-1 )) - \non
& & c(x_1, y)(t_d) \cdot [1-(1-c(x_1, y)(t-1))^8)] \}
\end{eqnarray}

\begin{eqnarray}
\label{evol6}
\Delta c_7(x, y_1) (t) & \equiv & \frac{1}{8} \cdot c(x, y)(t-1)  \{ ((1-c(x, y_1)(t_d)) \cdot (1- c^8(x, y_1) (t -1))- \non
& & c(x, y_1)(t_d) \cdot [1-(1-c(x, y_1)(t-1))^8)] \}
\end{eqnarray}

\begin{eqnarray}
\label{evol7}
\Delta c_8(x_2, y) (t) & \equiv & \frac{1}{8} \cdot c(x, y)(t-1)  \{ ((1-c(x_2, y)(t_d)) \cdot (1-c^8(x_2, y) (t-1) ) - \non
& &  c(x_2, y)(t_d) \cdot [1-(1-c(x_2, y)(t-1)^8)] \}
\end{eqnarray}

\begin{eqnarray}
\label{evol8}
\Delta c_9(x, y_2) (t) & \equiv & \frac{1}{8} \cdot c(x, y) (t-1) \{ ((1-c(x, y_2)(t_d) \cdot (1-c^8(x, y_2)(t-1)) - \non
& & c(x, y_2)(t_d) \cdot [1-(1-c(x, y_2)(t-1))^8)] \}
\end{eqnarray}

where 
$c(t_d)$ is the recently updated information about the system, 
$x_1 = x + J \cdot c(x,y)$, $x_2 = x - J \cdot c(x,y) $, $y_1 = y + J \cdot c(x,y) $, $y_2 = y- J \cdot c(x,y) $;each concentration is measured at the time indicated in Eqs. (\ref{evol1} - \ref{evol8}).\\

The new state of the system is calculated summing inputs from Eqs.(\ref{evol_new}-\ref{evol8}), i.e.,
\be
\label{evol_new_final}
c(x,y)(t) = c(x,y)(t-1) + \sum_{i=1}^9 \Delta c_i (x,y)(t) 
\ee

If the concentration of companies is smaller than zero,  a local crash of the system  is recognised  and the concentration is reset to zero. Unlike in the previously analysed system \cite{ACP_AM3} where after the crash the system has to remain in a zero state, here a local crash does not result in a total catastrophe. Due to the jumps it is possible that the place will be taken by a company coming  from a different lattice point later, or by one being created.

\section{Behaviour of the system}

\subsection{Time delay influence}

Examples of the system evolution are presented in Figs.\ref{plot_t_1}-\ref{plot_t_12}. The simulation parameters are as follows: the size of the square lattice $100 \times 100$. The jump length $J=4$, the time delay $t_d$ varies from 1 to 12. Initially the lattice was occupied in the middle $x\in (45,55)$, $y \in (45,55)$. The value of the initial concentration was chosen randomly from the interval $(0,1)$.

For the case with no delay in the information flow the system evolves toward a stable state with concentration $c=0.5$ at every lattice point. This is similar to the case discussed in the uniformly occupied system (Sec.\ref{uniform}), where the system evolved to the stable state in the case of short time delay of the information flow. However in the heterogeneous system an oscillatory behaviour becomes visible even for very short $t_d$. For $t_d = 2 \; IS$ the Fourier transform of the total concentration presented in Fig.\ref{fft_1} has a peak at $t=75 \; IS$. The same is for $t_d=4 \; IS$, but the maximum is significantly higher. It is in agreement with the evolution of the system presented in Fig.\ref{plot_t_4}, where oscillations of the total concentration can be seen. If the time delay is longer than 4 $IS$ the periodic behaviour becomes more complicated; the Fourier transform (Fig.\ref{fft_1}) shows two distinct maxima. The complexity of the periodic behaviour increases with the value of the time delay of information flow. For $t_d=8 \; IS$ there are three strong maxima and for $t_d=10 \; IS$ -- ten maxima.

\begin{figure}
	\centering
	\includegraphics[bb=50 50 410 302,scale=0.5]{miskiewicz_concentracja_x_1.eps}
	\includegraphics[bb=50 50 410 302,scale=0.5]{miskiewicz_concentracja_suma_1.eps}
\caption{ \label{plot_t_1}
Concentration evolution for a time delay $t_d=1 \; IS$; on the left: concentration summed up along Y axis, on the right the total concentration of the system}
\end{figure}

\begin{figure}
	\centering
	\includegraphics[bb=50 50 410 302,scale=0.5]{miskiewicz_concentracja_x_4.eps}
	\includegraphics[bb=50 50 410 302,scale=0.5]{miskiewicz_concentracja_suma_4.eps}
\caption{ \label{plot_t_4}
Concentration evolution for a time delay $t_d=4 \; IS$; on the left: concentration summed up along Y axis, on the right the total concentration of the system}
\end{figure}

\begin{figure}
	\centering
	\includegraphics[bb=50 50 410 302,scale=0.5]{miskiewicz_concentracja_x_8.eps}
	\includegraphics[bb=50 50 410 302,scale=0.5]{miskiewicz_concentracja_suma_8.eps}
\caption{ \label{plot_t_8}
Concentration evolution for a time delay $t_d=8 \; IS$; on the left: concentration summed up along y axis, on the right the total concentration of the system}
\end{figure}

\begin{figure}
	\centering
	\includegraphics[bb=50 50 410 302,scale=0.5]{miskiewicz_concentracja_x_12.eps}
	\includegraphics[bb=50 50 410 302,scale=0.5]{miskiewicz_concentracja_suma_12.eps}
\caption{ \label{plot_t_12}
Concentration evolution for a time delay $t_d=12 \; IS$; on the left: concentration summed up along Y axis, on the right the total concentration of the system}
\end{figure}

\begin{figure}
	\centering  
	\includegraphics[bb=50 50 410 302,scale=0.5]{miskiewicz_fft_1.eps}
	\includegraphics[bb=50 50 410 302,scale=0.5]{miskiewicz_fft_2.eps}
	\includegraphics[bb=50 50 410 302,scale=0.5]{miskiewicz_fft_4.eps}
	\includegraphics[bb=50 50 410 302,scale=0.5]{miskiewicz_fft_6.eps}
	\includegraphics[bb=50 50 410 302,scale=0.5]{miskiewicz_fft_8.eps}
	\includegraphics[bb=50 50 410 302,scale=0.5]{miskiewicz_fft_10.eps}
	\caption{\label{fft_1} Fourier transform of the total concentration for $t_d=1,2,4,6,8,10  \; IS$ as a function of the period length.}
	
\end{figure}

\subsection{Jump length influence}

The influence of the jump length parameter has been investigated in the case $J=10$, which is a significant number as compared with the lattice size, for various initial concentrations,. Only the speed of spreading of the companies over the system increased, e.g. for the case of no time delay the stable concentration of companies was achieved after 70 $IS$ ($J=4$) vs. 26 $IS$ ($ J=10$). Other results such as  the periodic behaviour or the stabilisation level did not change significantly.

\section{Conclusions}

The causes of economic cycles and their properties are fascinating problems in economy \cite{Kalecki1,Kalecki2,Kalecki3,others1,others2a,others2b,others2c,dresno}. In the present investigations it has been shown that the ability to move entities may prevent the system from a total crash. In contrast to the uniformly occupied system analysed in \cite{ACP_AM3} even in the case of a very long time delay of information flow, the systems do not collapse. On the other hand for $t_d > 4$ the system evolution reveals periodic behaviours. This may suggest that the information flow plays one of the key factors in economic cycle generations. It is worth to notice that $t_d= 4 \; IS$ causes periodic behaviour of $75 \; IS$. Applying the same proportion to real systems the quarterly updated data may result in a 5-6 year cycle. Of course the investigated model is highly far from any true complexity of any economic system, but the importance of the information flow in economy systems is hereby well demonstrated.

\section{Acknowledgements}
We would like to thank organisers of APFA 5 for their hospitality.
JM would like to thank EWFT for financial support allowing his participation in APFA 5 conference.

\bibliographystyle{unsrt} 
\bibliography{miskiewicz}

\begin{thebibliography}{10}

\bibitem{games1}
M.~J. Osborne and A.~Rubinstein.
\newblock {\em A course in game theory by Martin}.
\newblock MIT Press, 1994.

\bibitem{games2}
A.~V. Banerjee.
\newblock A simple model of herd behavior.
\newblock {\em The Quarterly Journal of Economics}, 107(3):797--817, August
  1992.
\newblock available at
  http://ideas.repec.org/a/tpr/qjecon/v107y1992i3p797-817.html.

\bibitem{games3}
L.~M. Ausubel, P.~Cramton, and R.~J. Deneckere.
\newblock Bargaining with incomplete information.
\newblock Papers of Peter Cramton 02barg, University of Maryland, Department of
  Economics - Peter Cramton, 2002.
\newblock available at http://ideas.repec.org/p/pcc/pccumd/02barg.html.

\bibitem{MA1}
B.~Finkenst\"adt.
\newblock {\em Nonlinear Dynamics in Economics: A theoretical and statistical
  approach to agricultural markets}.
\newblock Number 426 in Springer Lecture Notes in Economics and Mathematical
  Systems. Springer, Berlin, 1995.

\bibitem{vitanov}
N.~Vitanov.
\newblock private comunication.

\bibitem{ACP3}
M.~Ausloos, P.~Clippe, and A.~P\c{e}kalski.
\newblock Evolution of economic entities under heterogenous
  political/environmental conditions within bak-sneppen-like dynamics.
\newblock {\em Physica A}, 332:394--402, 2004.

\bibitem{ACP2}
M.~Ausloos, P.~Clippe, and A.~P\c{e}kalski.
\newblock Model of macroeconomic evolution in stable regionally dependent
  economic fields.
\newblock {\em Physica A}, 337:269--287, 2004.

\bibitem{ACP1}
M.~Ausloos, P.~Clippe, and A.~P\c{e}kalski.
\newblock Simple model for the dynamics of correlation in the evolution of
  economic entities under varying economic conditions.
\newblock {\em Physica A}, 324:330--337, 2003.

\bibitem{ACP_AM3}
J.~Mi\'skiewicz and M.~Ausloos.
\newblock Influence of information flow in the formation of economic cycles.
\newblock In m.~Ausloos and M.~Dirickx, editors, {\em The Logistic Map and the
  Route to Chaos}, Understanding Complex Systems, pages 223 -- 238.
  Springer-Verlag Berlin Heildelbrg, 2006.

\bibitem{ACP_AM1}
J.~Mi\'skiewicz and M.~Ausloos.
\newblock A logistic map approach to economic cycles. (i) the best adapted
  companies.
\newblock {\em Physica A}, 336:206--214, 2004.

\bibitem{ACP_AM2}
M.~Ausloos, P.~Clippe, J.~Mi\'skiewicz, and A.~P\c{e}kalski.
\newblock A (reactive) lattice-gas approach to economic cycles.
\newblock {\em Physica A}, 344:1--7, 2004.

\bibitem{Kalecki1}
M.~Kalecki.
\newblock A macrodynamic theory of business cycles.
\newblock {\em Econometrica}, 3:327, 1935.

\bibitem{Kalecki2}
M.~Kalecki.
\newblock A theory of the business cycle.
\newblock {\em Rev. Econ. Studies}, 4:77, 1937.

\bibitem{Kalecki3}
M.~Kalecki.
\newblock {\em Theory of Economic Dynamics: An essay on cyclical and long-run
  changes in capitalist economy}.
\newblock Monthly Review Press, New York, 1965.

\bibitem{others1}
G.~Gabisch and H.W. Lorenz.
\newblock {\em Business Cycle Theory: A survey of methods and concepts}.
\newblock Springer-Verlag, Berlin, 1989.

\bibitem{others2a}
M.~Aoki.
\newblock Asymmetrical cycles and equilibrium selection in finitary
  evolutionary economic models.
\newblock In L.~Punzo, editor, {\em Cycles, Growth, and Structural Changes},
  chapter~8. Routledge, London, 2001.

\bibitem{others2b}
M.~Aoki.
\newblock A simple model of asymmetrical business cycles: Interactive dynamics
  of a large number of agents with discrete choices.
\newblock {\em Macroeconomic Dynamics}, 2:427, 1998.

\bibitem{others2c}
M.~Aoki.
\newblock Stochastic views on diamond search model: Asymmetrical cycles and
  fluctuations.
\newblock {\em Macroeconomic Dynamics}, 4:487, 2000.

\bibitem{dresno}
H.~G. Danielmayer.
\newblock On the nature of business cycles.
\newblock In {\em European Conference Abstracts}, volume 30A, page 686, 2006.

\end{thebibliography}

\end{document}